\documentclass{PoS}

\usepackage[utf8x]{inputenc}
\usepackage{amsmath}
\usepackage{mathtools}
\usepackage{nicefrac}
\usepackage{bm}
\usepackage{cite}
\usepackage{float}
\usepackage[merge,sort&compress,numbers]{natbib}
\title{London penetration depth and coherence length of SU(3) vacuum flux
tubes}

\ShortTitle{London penetration depth and coherence length of SU(3) vacuum flux
tubes}

\author{Paolo Cea\\
        Dipartimento di Fisica dell'Università di Bari \\and INFN - Sezione di Bari, I-70126 Bari, Italy\\
        E-mail: \email{paolo.cea@ba.infn.it}}
\author{Leonardo Cosmai\\
        INFN - Sezione di Bari, I-70126 Bari, Italy\\
        E-mail: \email{leonardo.cosmai@ba.infn.it}}
\author{\speaker{Francesca Cuteri}\\
        Dipartimento di Fisica dell'Università della Calabria \\and INFN - Gruppo collegato di Cosenza, I-87036 Arcavacata di Rende, Cosenza, Italy\\
        E-mail: \email{francesca.cuteri@fis.unical.it}}
\author{Alessandro Papa\\
        Dipartimento di Fisica dell'Università della Calabria \\and INFN - Gruppo collegato di Cosenza, I-87036 Arcavacata di Rende, Cosenza, Italy\\
        E-mail: \email{alessandro.papa@fis.unical.it}}

\abstract{The transverse profile of the chromoelectric field generated by a quark-
antiquark pair in the SU(3) vacuum is analysed within the dual superconductor
scenario, then the London penetration depth and coherence length are extracted.
The color field is determined on the lattice through a connected correlator of
two Polyakov loops measured on smeared configurations.}

\FullConference{The 32nd International Symposium on Lattice Field Theory,\\
		23-28 June, 2014\\
		Columbia University New York, NY}

\begin{document}

\section{Introduction}
As is well known, the chromoelectric flux tubes produced by a pair of static 
color charges in the QCD vacuum represent an evidence for the confinement
phenomenon~\cite{Bander:1980mu,*Greensite:2003bk}. Monte Carlo simulations of
lattice QCD~\cite{Fukugita:1983du,*Kiskis:1984ru,*Flower:1985gs,*Wosiek:1987kx,
*Singh:1993jj,*Cea:1992sd,*Bali:1994de,*Haymaker:2005py,*D'Alessandro:2006ug,
DiGiacomo:1990hc,Cea:1992vx,*Cea:1993pi,*Cea:1994ed,*Cea:1994aj} allow
nonperturbative studies of the chromoelectric field distribution associated with
the flux-tube structures. Within the dual superconductor model of QCD vacuum,
conjectured by 't Hooft and Mandelstam~\cite{Mandelstam:1974pi,*'tHooft:1976ep,
*Ripka:2003vv}, the condensation of color magnetic monopoles responsible for the
formation of flux tubes is seen in analogy to the formation of Cooper pairs in the
BCS theory of superconductivity. Whereas the dynamical condensation of color
magnetic monopoles is not explained by the dual superconductor construction,
convincing lattice evidences for this condensation mechanism have been
found~\cite{Shiba:1994db,*Arasaki:1996sm,*Cea:2000zr,*Cea:2001an,
*DiGiacomo:1999fb,*Carmona:2001ja,*Cea:2004ux,*D'Alessandro:2010xg}. 
In previous studies~\cite{Cea:1992vx,*Cea:1993pi,*Cea:1994ed,*Cea:1994aj,
Cea:1995zt,*Cardaci:2010tb}, the flux-tube chromoelectric field distribution has
been investigated through the connected correlation
function~\cite{DiGiacomo:1990hc,DiGiacomo:1989yp,*Kuzmenko:2000bq,
*DiGiacomo:2000va}:
\begin{equation}
\label{rhoW}
\rho_W^{\rm conn} = \frac{\left\langle {\rm tr}
\left( W L U_P L^{\dagger} \right)  \right\rangle}
              { \left\langle {\rm tr} (W) \right\rangle }
 - \frac{1}{N} \,
\frac{\left\langle {\rm tr} (U_P) {\rm tr} (W)  \right\rangle}
              { \left\langle {\rm tr} (W) \right\rangle } \; ,
\end{equation}
where $U_P=U_{\mu\nu}(x)$ is the plaquette in the $(\mu,\nu)$ plane, connected to
the Wilson loop $W$ by a Schwinger line $L$, and $N$ is the number of colors (see
Fig.~1 in Refs.~\cite{Cea:1995zt,*Cardaci:2010tb}). In the naive continuum
limit~\cite{DiGiacomo:1990hc} we have
\begin{equation}
\label{rhoWlimcont}
\rho_W^{\rm conn}\stackrel{a \rightarrow 0}{\longrightarrow} a^2 g 
\left[ \left\langle
F_{\mu\nu}\right\rangle_{q\bar{q}} - \left\langle F_{\mu\nu}
\right\rangle_0 \right]  \;,
\; \; \; \; \; \; \; \;\; \; \; \; \; \; \; \;F_{\mu\nu}(x) = \sqrt\frac{\beta}{2 N} \, \rho_W^{\rm conn}(x)   \; .
\end{equation}
where $\langle\quad\rangle_{q \bar q}$ denotes the average in the presence of a
static $q \bar q$ pair and $\langle\quad\rangle_0$ is the vacuum average.

In ordinary superconductivity tube-like structures arise as solutions of the
Ginzburg-Landau equations~\cite{Abrikosov:1957aa}.
Within dual superconductivity, the formation of the chromoelectric flux tubes can
be interpreted as dual Meissner effect and the chromoelectric field distribution
should resemble the dual version of the Abrikosov vortex field distribution. This
led to the proposal~\cite{Cea:1992vx,Cea:1993pi,Cea:1994ed,Cea:1994aj,Cea:1995zt}
to fit the transverse shape of the longitudinal chromoelectric field according to
\begin{equation}
\label{London}
E_l(x_t) = \frac{\Phi}{2 \pi} \mu^2 K_0(\mu x_t) \;,\;\;\;\;\; x_t > 0 \; ,
\end{equation}
where $K_n$ is the modified Bessel function of order $n$, $\Phi$ is the external
flux, and $\lambda=1/\mu$ is the London penetration length. However,
Eq.~(\ref{London}) is valid only for type-II superconductors, i.e. for
$\lambda \gg \xi$, $\xi$ being the coherence length, which measures the coherence
of the magnetic monopole condensate.
Several numerical studies~\cite{Suzuki:1988yq,*Maedan:1989ju,*Singh:1992ma,
*Matsubara:1993nq,*Schlichter:1997hw,*Bali:1997cp,*Schilling:1998gz,
*Gubarev:1999yp,*Koma:2003hv} have, instead, shown that the
confining vacuum behaves much like a dual superconductor lying on the borderline
between type-I and type-II superconductivity.
Nonetheless, in Ref.~\cite{Cea:2012qw} it has been suggested a different fitting
function by exploiting the results in Ref.~\cite{Clem:1975aa}. There, from the
assumption of a simple variational model for the magnitude of the normalized order
parameter of an isolated vortex, analytic expressions for magnetic field and
supercurrent density are derived, that solve the Ampere's law and the
Ginzburg-Landau equations. By dual analogy
\begin{equation}
\label{clem1}
E_l(x_t) = \frac{\phi}{2 \pi} \frac{1}{\lambda \xi_v} \frac{K_0(R/\lambda)}
{K_1(\xi_v/\lambda)} \,,\; \; \; \; \; \; \; \;  R=\sqrt{x_t^2+\xi_v^2}  \;,
\end{equation}
where $\xi_v$ is a variational core-radius parameter. Equation~(\ref{clem1})
is equivalent to
\begin{equation}
\label{clem2}
E_l(x_t) =  \frac{\phi}{2 \pi} \frac{\mu^2}{\alpha} \frac{K_0[(\mu^2 x_t^2 
+ \alpha^2)^{1/2}]}{K_1[\alpha]} \; ,
\; \; \; \; \; \; \; \;\mu= \frac{1}{\lambda} \,, \quad \frac{1}{\alpha} =
\frac{\lambda}{\xi_v} \,.
\end{equation}
By fitting Eq.~(\ref{clem2}) to $E_l(x_t)$ data, one can extract both the
penetration length $\lambda$ and $\lambda/\xi_v$. The Ginzburg-Landau $\kappa$
parameter can then be obtained by
\begin{equation}
\label{landaukappa}
\kappa = \frac{\lambda}{\xi} =  \frac{\sqrt{2}}{\alpha} 
\left[ 1 - K_0^2(\alpha) / K_1^2(\alpha) \right]^{1/2} \,,
\end{equation}
and, the coherence length $\xi$ can be deduced.

\begin{figure}[h] 
\label{Loop}
\centering
\includegraphics[width=0.45\textwidth]{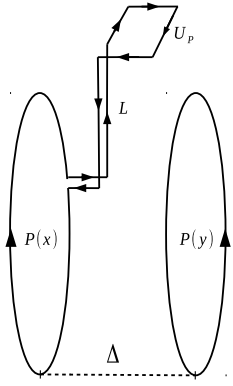} 
\caption{The connected correlator given in Eq.~(\protect\ref{eq:rhopconn})
(subtraction in $\rho_{P}^{\rm conn}$ not explicitly drawn).}
\end{figure}

With the final aim of extending the analysis of flux tubes to the case of finite
temperatures (where the study of these structures is directly relevant to clarify
the formation of $c \bar{c}$ and $b \bar{b}$ bound states in heavy ion collisions),
we cannot employ Eq.~(\ref{rhoW}). Nevertheless, it suffices replacing, in
Eq.~(\ref{rhoW}), the Wilson loop with two Polyakov lines (see Fig.~\ref{Loop}).
In addition, also the cooling mechanism, previously used to enhance the
signal-to-noise ratio, was replaced in our work by the APE smearing
procedure~\cite{Falcioni:1984ei,*Albanese:1987ds}, to get rid of lattice artifacts.
Preliminarily, a check that this method gives results consistent with previous
studies, adopting Wilson loops and cooling, is necessary, and that is the subject
of the present work (see Ref.~\cite{Cea:2014uja} for more details).
Indeed, numerical results on the chromoelectric flux tubes in SU(3) pure gauge
theory at zero temperature, obtained with connected correlations built with
Polyakov lines and smeared gauge links, are presented in what follows.
%

\section{Flux tubes on the lattice}
In order to explore the field configurations produced by a static $q\overline{q}$
pair, the following connected correlation function was considered:
\begin{equation}
\label{eq:rhopconn}
\rho_{P}^{\rm conn} = \frac{\left\langle \mathrm{tr}\left(P\left(x\right)LU_{P}
L^{\dagger}\right)\mathrm{tr}P\left(y\right)\right\rangle }{\left\langle 
\mathrm{tr}\left(P\left(x\right)\right)\mathrm{tr}\left(P\left(y\right)\right)
\right\rangle } - \frac{1}{3}\frac{\left\langle \mathrm{tr}\left(P\left(x\right)
\right)
\mathrm{tr}\left(P\left(y\right)\right)\mathrm{tr}\left(U_{P}\right)\right
\rangle }{\left\langle \mathrm{tr}\left(P\left(x\right)\right)\mathrm{tr}
\left(P\left(y\right)\right)\right\rangle}\;
\end{equation}
The two Polyakov lines are separated by a distance $\Delta$. In the continuum
limit we obtain the field strength tensor, defined as~\cite{Skala:1996ar} 
\begin{equation}
\label{fieldstrengthP}
 F_{\mu\nu}\left(x\right)=\sqrt{\frac{\beta}{6}}\rho_{P}^{\rm conn}
\left(x\right).
\end{equation}
\begin{figure}[h]
\label{campo} 
  \centering
   \includegraphics[width=0.46\linewidth]{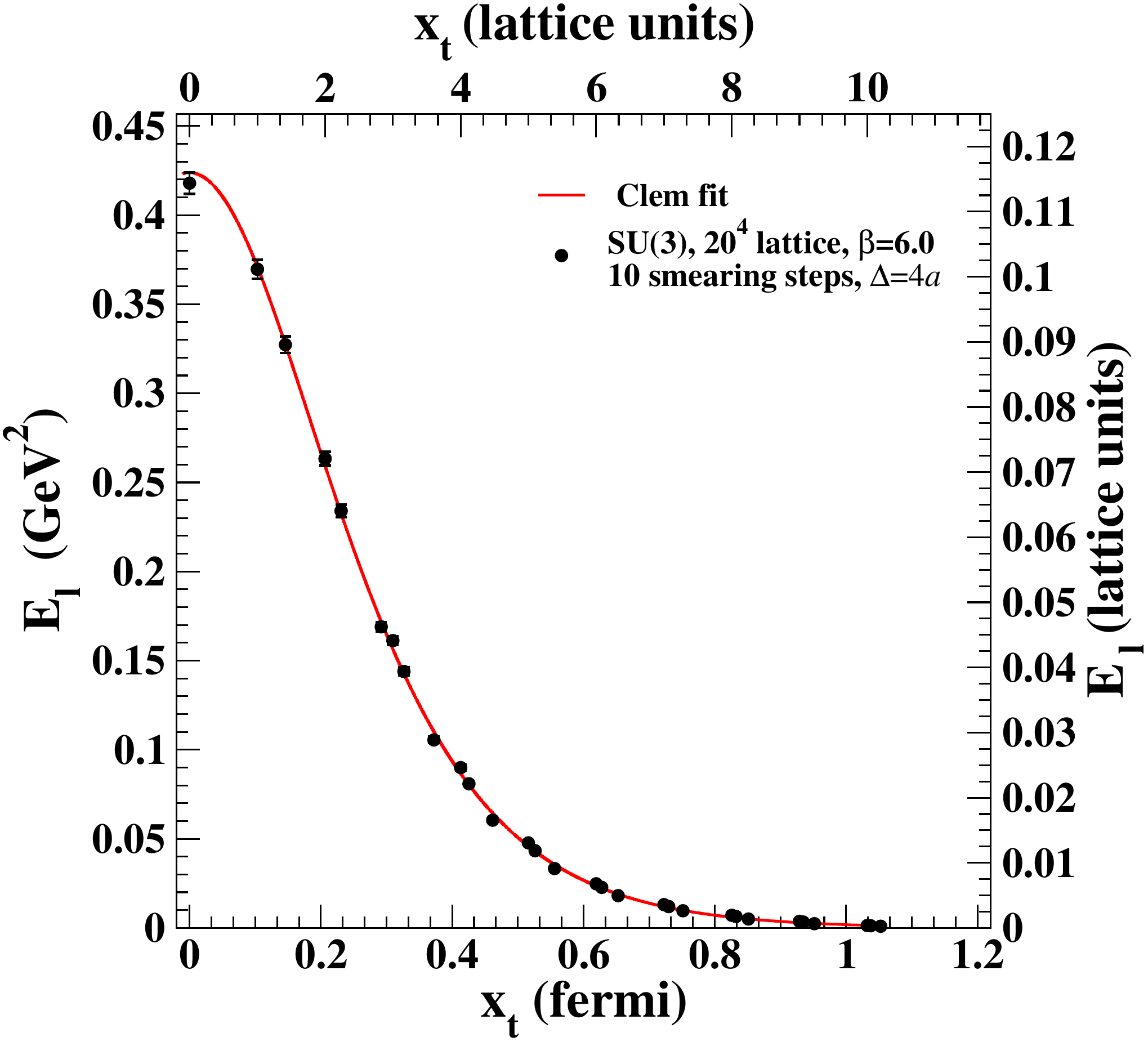} 
   \includegraphics[width=0.46\linewidth]{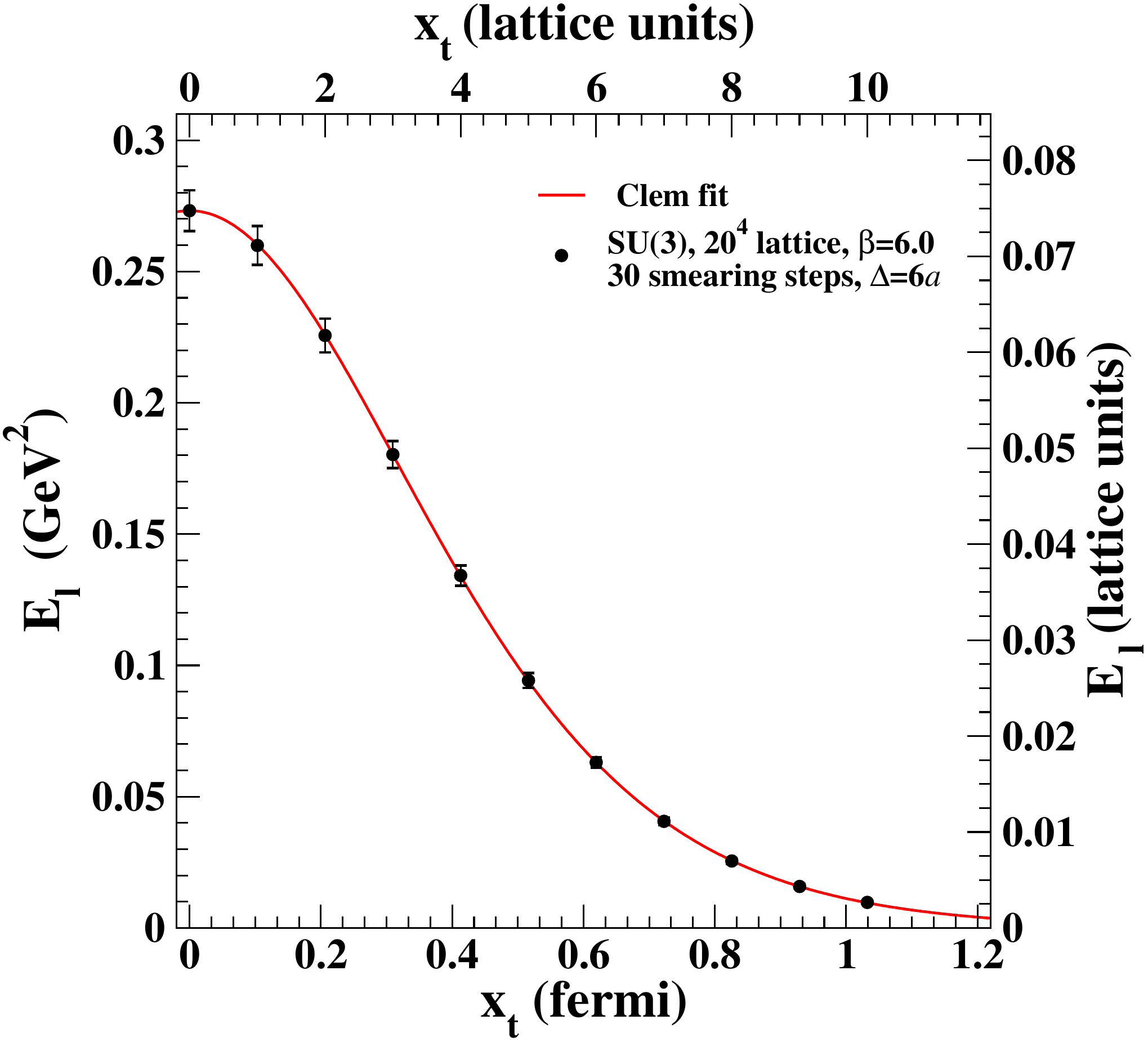}
   \scriptsize{\caption{ (Left) $E_{l}$ versus $x_t$, in lattice units and in
            physical units, at $\beta=6.0$ and for $\Delta=4a$, after 10 smearing
            steps. Full line is the best fit using Eq.~(\protect\ref{clem2}).
            (Right) The same, but for $\Delta=6a$, after 30 smearing steps. The
            procedure to fix the physical scale is explained in Sect.~3.}}
\end{figure}
Wilson action with periodic boundary conditions and the Cabibbo-Marinari 
algorithm~\cite{Cabibbo:1982zn} combined with overrelaxation were used,
and simulations were performed on $20^4$ lattices.
We considered $\Delta = 4a, 6a, 8a$ ($a$ is the lattice spacing), and four
different values of the gauge coupling $\beta$ in the range $5.9 \div 6.1$.
Measurements were taken every 10 updatings in order to reduce the autocorrelation
time. The jackknife method was used for the error analysis. The smearing procedure
as described in Ref.~\cite{Falcioni:1984ei,*Albanese:1987ds}, with smearing
parameter $\epsilon = 0.5$, was employed to reduce statistical errors.
The flux tube is confirmed to be almost completely formed by the longitudinal
chromoelectric field $E_l$, which is constant along the flux axis and decreases
rapidly in the transverse direction $x_t$.
To probe $E_l(x_t)$, the plaquette in Eq.~(\ref{clem2}) was placed in
correspondence to the middle point (labeled by $x_t=0$) of the axis connecting
the static sources and, then, moved along all possible transverse spatial
directions to distances $x_t>0$ from that axis.
We fitted our data to Eq.~(\ref{clem2}) and the result of the fit is shown in
Fig.~\ref{campo}. 
To check rotational invariance, also noninteger distances were considered for
$\Delta = 4a$, but, since the only effect of restricting the fit to integer
distances was a reduction of the reduced chi-square, $\chi_r^2$, in order to have
less time-consuming simulations, we performed measurements for integer transverse
distances only, for all the other $\Delta$ values.
\begin{table}[tb]
\begin{center} 
\footnotesize{\caption{Summary of the fit values at $\beta=6.0$ for $\Delta=6a$.
\label{table:6.0_s6}}}
\scriptsize {\begin{tabular}{cccccc}
\hline\hline
Smearing& $\phi$ & $\mu$ & $\lambda/\xi_v$ & $\kappa$ & $\chi_r^2$ \\ \hline
 16&	6.191(141) &	0.621(79) &	0.309(95) &	0.213(91) &	0.018\\
 18&	6.218(125) &	0.622(76) &	0.287(82) &	0.192(77) &	0.011\\
 20&	6.227(109) &	0.617(68) &	0.277(72) &	0.183(66) &	0.010\\
 22&	6.222(98) &	0.608(61) &	0.271(64) &	0.178(58) &	0.010\\
 24&	6.207(88) &	0.597(55) &	0.269(58) &	0.176(53) &	0.011\\
 26&	6.184(81) &	0.587(50) &	0.269(54) &	0.175(49) &	0.011\\
 28&	6.155(75) &	0.576(47) &	0.269(51) &	0.176(46) &	0.011\\
 30&	6.122(70) &	0.566(44) &	0.270(48) &	0.176(44) &	0.010\\
 32&	6.087(66) &	0.557(41) &	0.271(46) &	0.177(42) &	0.009\\
 34&	6.049(63) &	0.549(39) &	0.271(45) &	0.178(41) &	0.008\\
 36&	6.011(60) &	0.541(37) &	0.272(43) &	0.179(40) &	0.007\\
 38&	5.973(58) &	0.534(36) &	0.273(42) &	0.179(39) &	0.005\\
 40&	5.935(56) &	0.527(35) &	0.274(42) &	0.180(38) &	0.004\\
 42&	5.897(54) &	0.521(34) &	0.274(41) &	0.180(37) &	0.003\\
 44&	5.859(53) &	0.515(33) &	0.275(40) &	0.181(37) &	0.003\\
 46&	5.822(51) &	0.510(32) &	0.275(40) &	0.181(37) &	0.002\\
 48&	5.786(50) &	0.505(31) &	0.276(39) &	0.182(36) &	0.002\\
 50&	5.751(49) &	0.500(31) &	0.277(39) &	0.182(36) &	0.001\\
\hline\hline 
\end{tabular} }
\end{center}
\end{table}
\begin{figure}[tbh]
  \centering
   \includegraphics[width=0.39\linewidth]{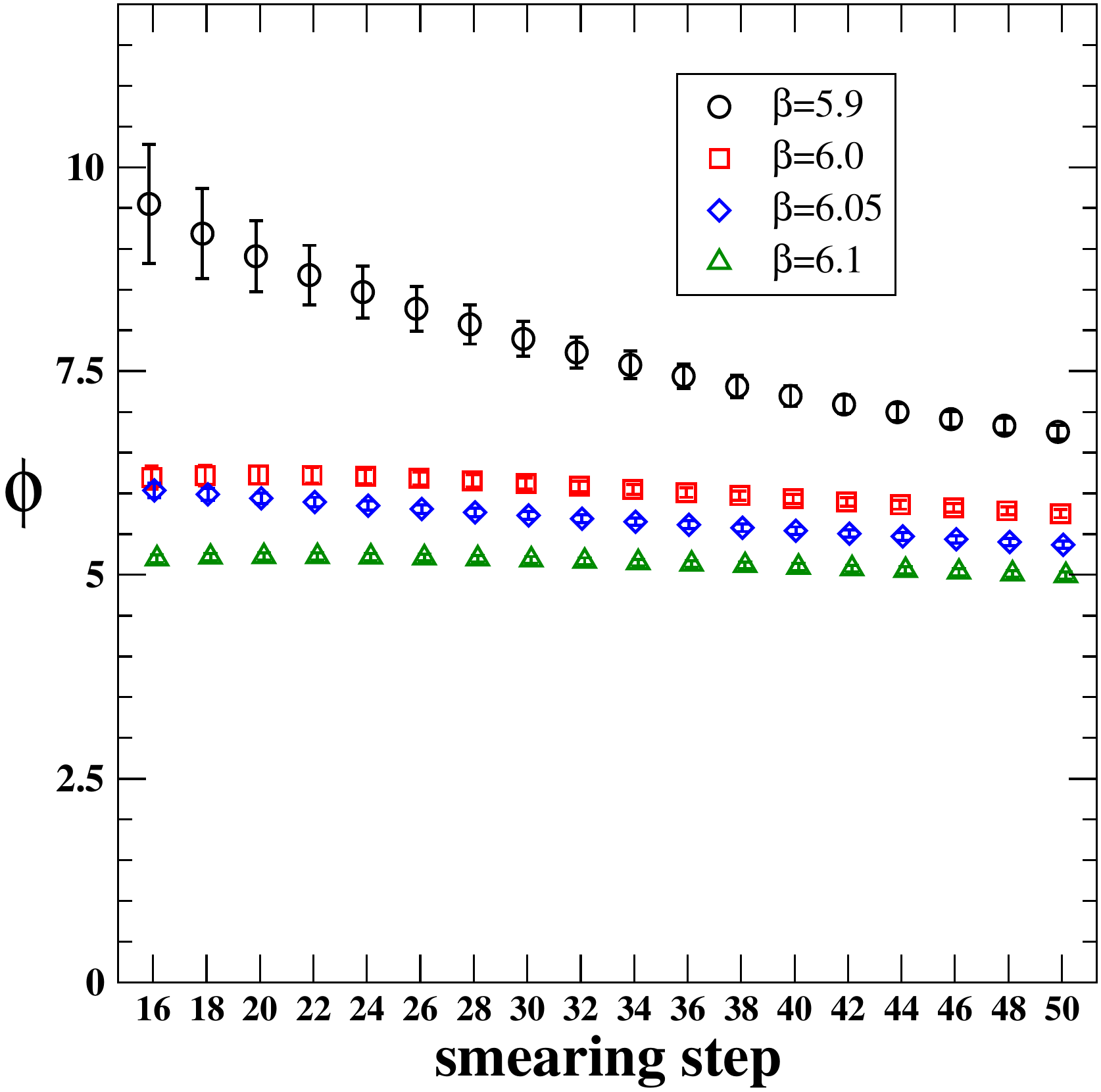}
   \includegraphics[width=0.4\linewidth]{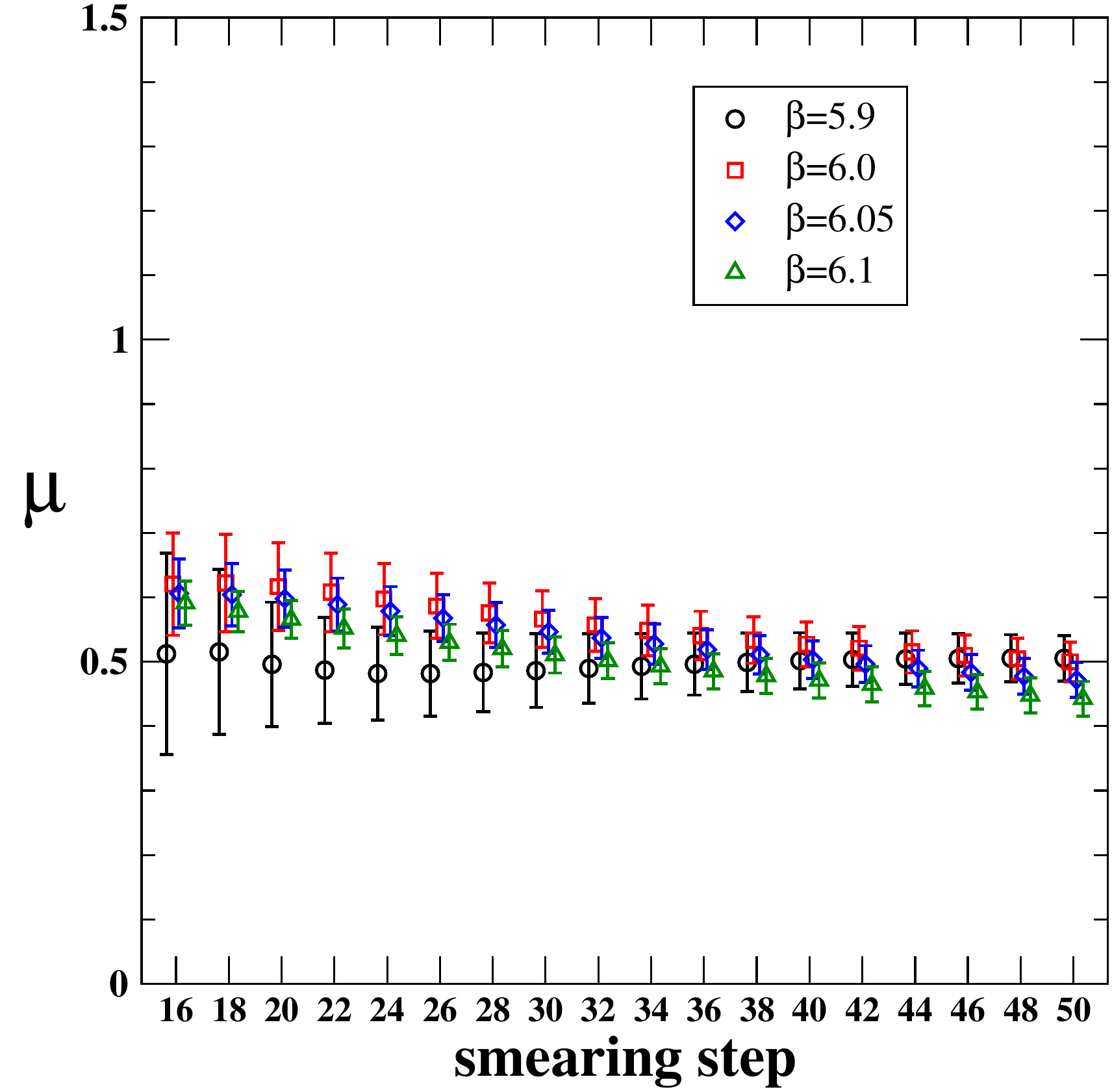}
    \rightskip=0.7in \includegraphics[width=0.41\linewidth]{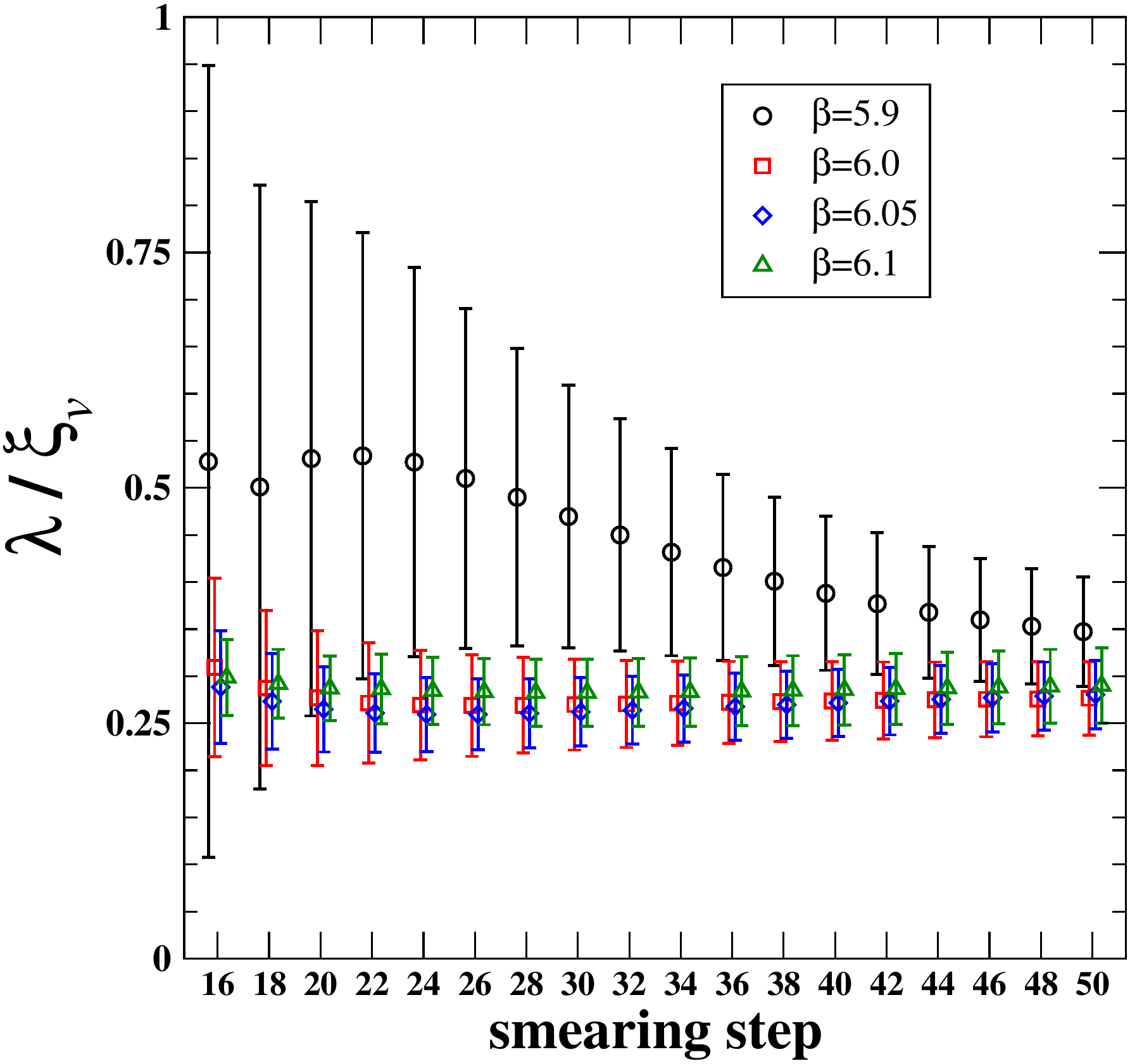}
   \includegraphics[width=0.4\linewidth]{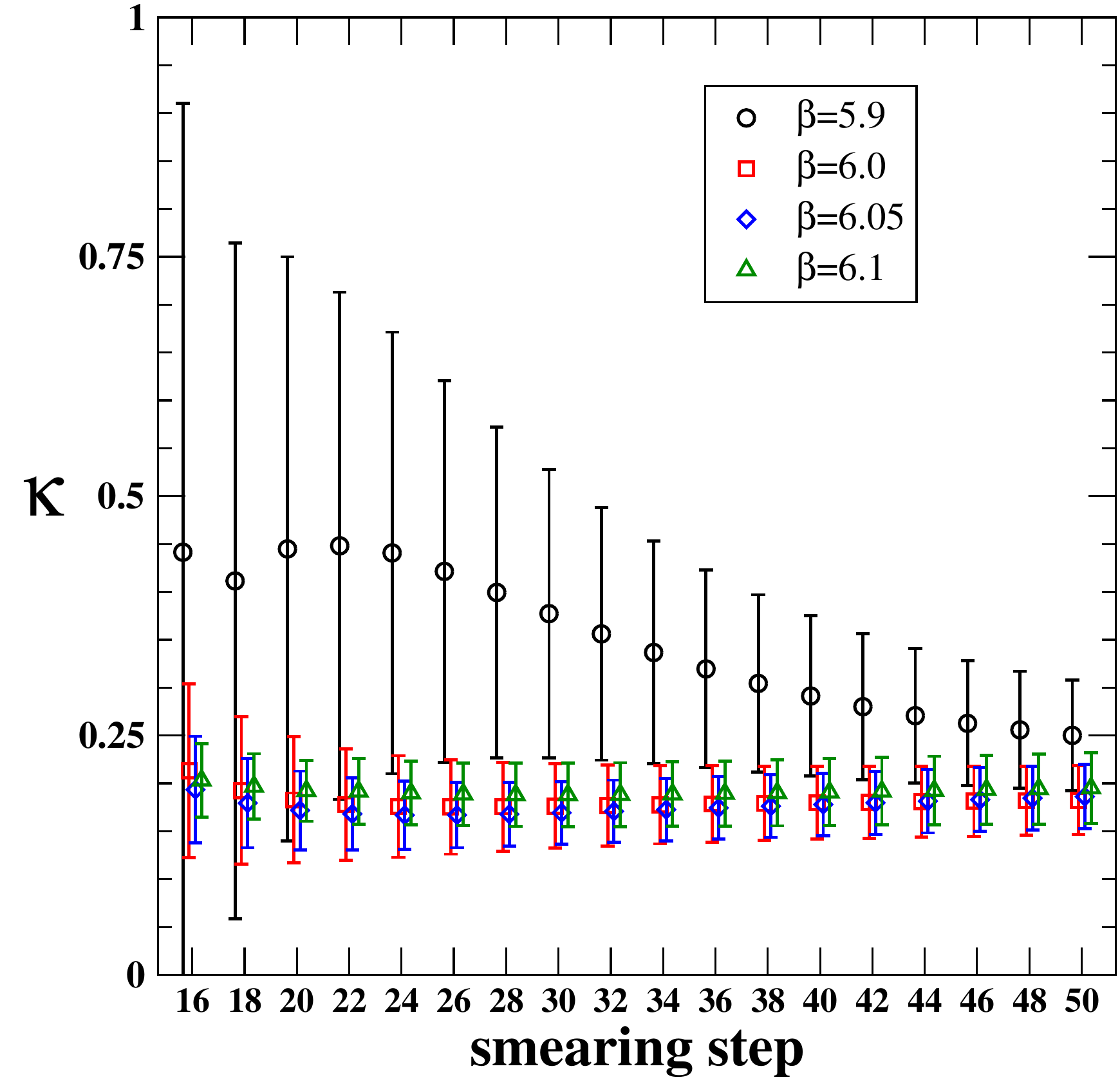} 
   \scriptsize{\caption{(Up left) $\phi$ versus the smearing step.
                        (Up right) The same for the inverse of the penetration length $\mu$.
                        (Down left) The same for $\nicefrac{\lambda}{\xi_v}$.
                        (Down right) The same for the Ginzburg-Landau parameter $\kappa$.
                        In all cases $\Delta=6a$, and in the last three figures data have
                        been slightly shifted along the horizontal axis for the sake of
                        readability.\label{Par.vs.smearing}}}
\end{figure}
%

The fit of our data to Eq.~(\ref{clem2}) was realized for each smearing step in
the interval $16 \div 50$. The parameters $\phi$, $\mu$,
and $\nicefrac{\lambda}{\xi_v}$ were extracted and the Ginzburg-Landau parameter
$\kappa$ was evaluated through Eq.~(\ref{landaukappa}).
Well-defined plateaux were found in the dependence of all parameters on the number
of smearing steps (see table~\ref{table:6.0_s6} for the $\beta$=6.0 and
$\Delta=6a$ case).
In order to check for contamination effects, on $E_l$, due to the proximity of
the static color sources, we varied $\Delta$. It was found that the $\mu$ and
$\lambda/\xi_v$ values for $\Delta = 4a$ were systematically higher than for
$\Delta = 6a, 8a$, while, for all the parameters the values obtained for the
$\Delta = 6a$ and $8a$ were consistent within each other.
The large statistical errors affecting our estimates for $\Delta = 8a$ led us to
focus our analysis on the case $\Delta = 6a$ as a good compromise between the
absence of contamination effects and a reasonable signal-to-noise ratio.
Figure~\ref{Par.vs.smearing} shows the behavior of $\phi$, $\mu$, $\lambda/\xi_v$
and $\kappa$, for $\Delta = 6a$ and for different values of $\beta$, versus the
number of smearing step. For $\beta \ge 6.0$, our estimate for the parameters
seems to be independent of the number of smearing steps.

\begin{figure}[h] 
  \centering
   \includegraphics[width=0.4\linewidth]{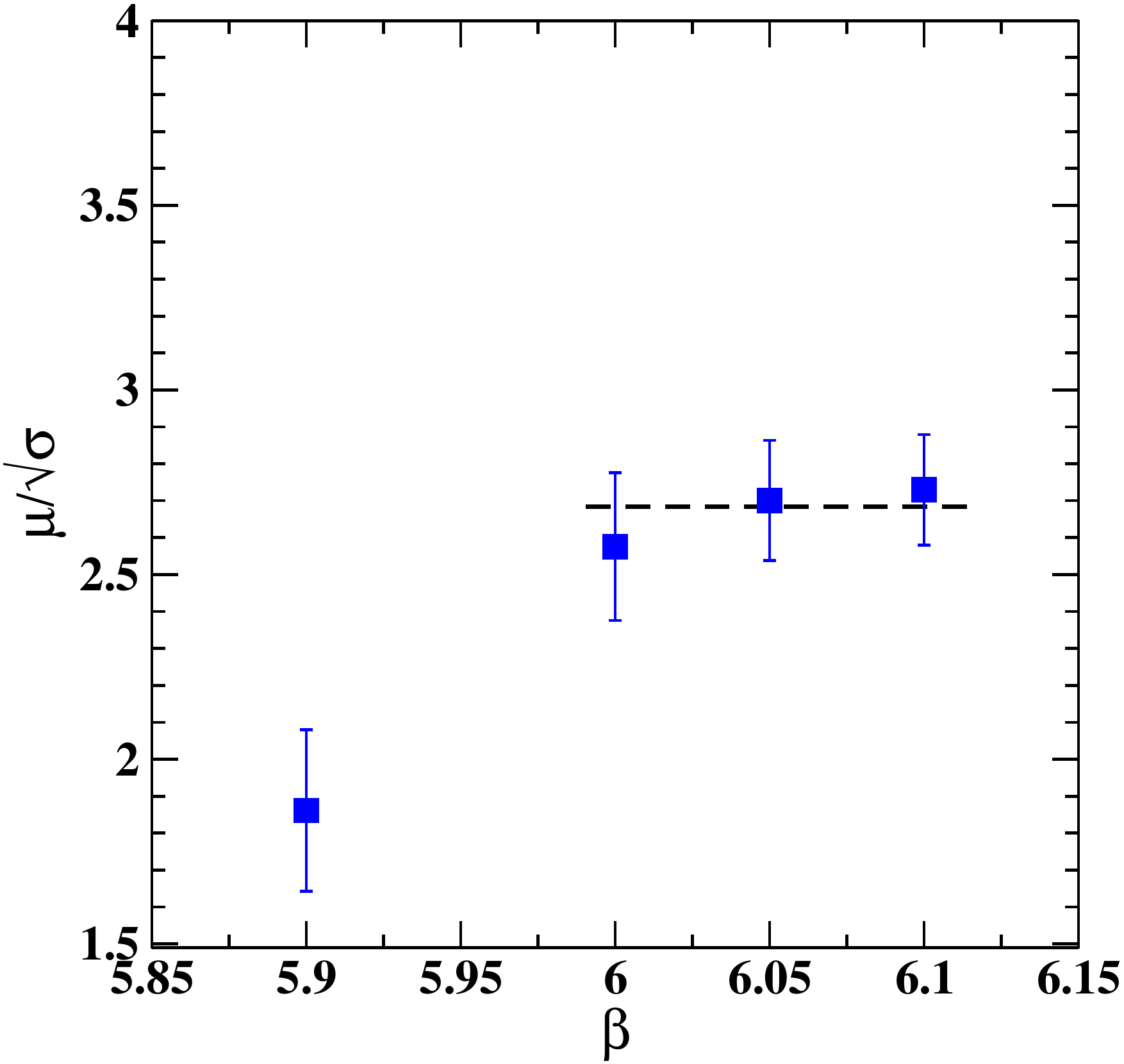}
   \includegraphics[width=0.4\linewidth]{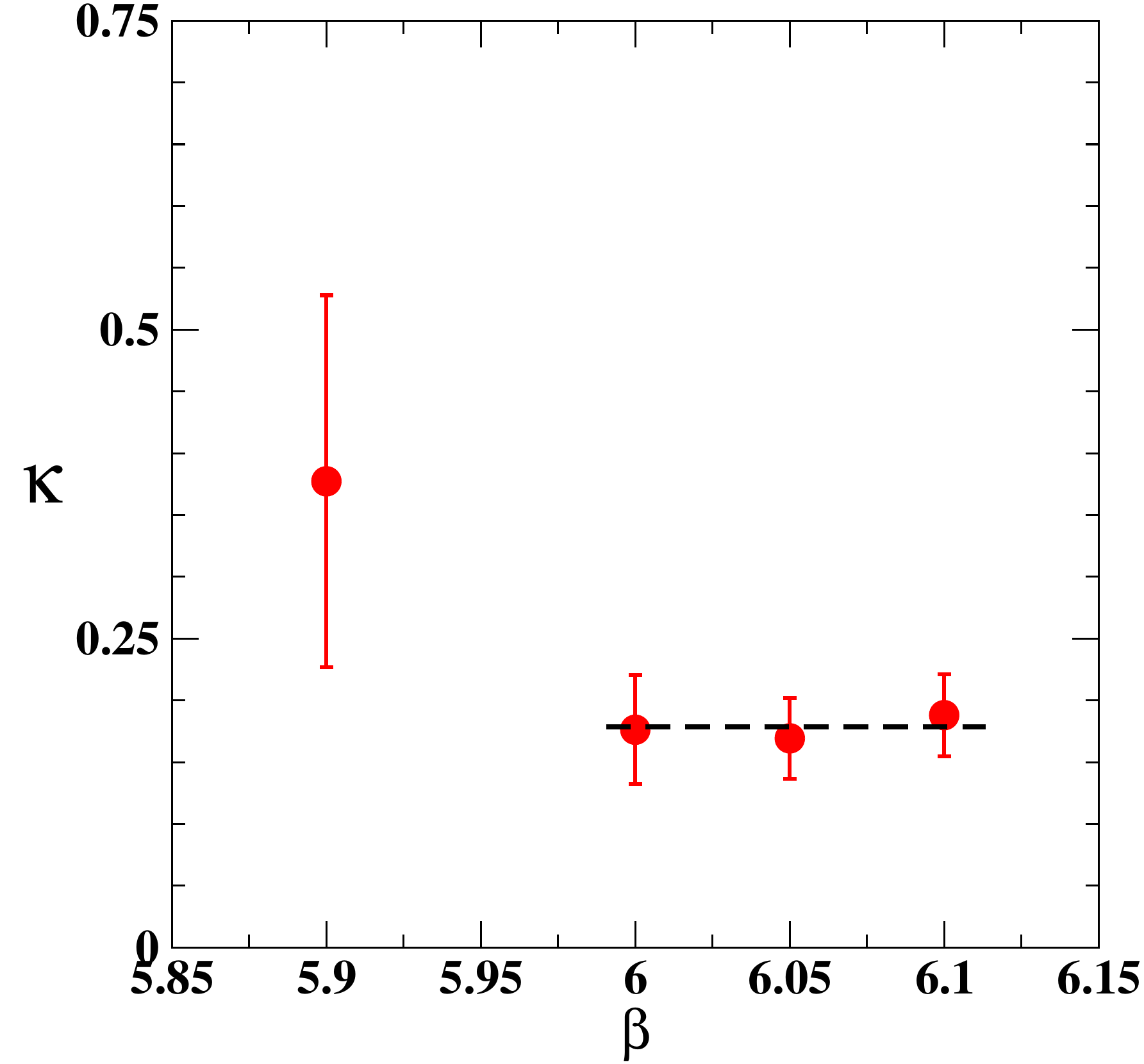}
  \scriptsize{\caption{(Left) $\mu/\sqrt{\sigma}$ versus $\beta$ for $\Delta=6a$.
                       (Right) $\kappa$ versus $\beta$ for $\Delta=6a$.\label{scaling}}}
\end{figure}
%

\section{Penetration and coherence lengths}
The estimation in physical units of the London penetration depth, $\lambda$,
and the coherence length, $\xi$, is the final goal of our analysis.
For this purpose it is, first of all, necessary to study the scaling of the
plateau values of $a\mu$ with the string tension. To do this, we expressed
the values of $a\mu$ in units of $\sqrt\sigma$, using the
parameterization~\cite{Edwards:1997xf}: 
\begin{equation}
\label{sqrt-sigma-SU3}
\sqrt{\sigma}(g)=f_{{\rm{SU(3)}}}(g^2)[1+0.2731\,\hat{a}^2(g)-0.01545\,\hat{a}^4(g) +0.01975\,\hat{a}^6(g)]/0.01364 \;,
\end{equation}
\[
\hat{a}(g) = \frac{f_{{\rm{SU(3)}}}(g^2)}{f_{{\rm{SU(3)}}}(g^2(\beta=6))} 
\;, \;
\beta=\frac{6}{g^2} \,, \;\;\; 5.6 \leq \beta \leq 6.5\;,
\]
where
\begin{equation}
\label{fsun}
f_{{\rm{SU(3)}}}(g^2) = \left( {b_0 g^2}\right)^{- b_1/2b_0^2} 
\, \exp \left( - \frac{1}{2 b_0 g^2}\right) \,,\; \; \; \; \; \; \; \; b_0 \, =
\, \frac{11}{(4\pi)^2} \; \; , \; \; b_1 \, = \, \frac{102}{(4\pi)^4} 
\; .
\end{equation}
The use of the above parameterization allowed us also to compute and display,
in Fig.~\ref{campo}, the transverse structure of $E_l(x_t)$ in physical units.
Figure~\ref{scaling} (Left) shows the ratio $\mu/\sqrt{\sigma}$ versus $\beta$.
For $\beta \ge 6.0$, $\mu$ scales according to the string tension.
Likewise, the dimensionless Ginzburg-Landau parameter $\kappa$ scales in the 
same interval of $\beta$ (see Fig.~\ref{scaling} (Right)).
Fitting, in both cases, the data in the scaling window with a constant we get
\begin{equation}
\label{mu_sqrt-sigma-SU3-&-kappa-phys}
\frac{\mu}{\sqrt{\sigma}} = 2.684(97) \,,\; \; \; \; \; \; \; \; \kappa \; = \;
0.178(21) \;.
\end{equation}
Assuming the standard value for the string tension, $\sqrt{\sigma}=420$~MeV, 
from Eq.~(\ref{mu_sqrt-sigma-SU3-&-kappa-phys}) we get
\begin{equation}
\label{lambda-phys}
\lambda \; = \; \frac{1}{\mu} \;  =  \; 0.1750(63) \; {\rm fm}  \;,
\; \; \; \; \; \; \; \; \xi  \;  =  \; 0.983(121) \; {\rm fm}  \; .
\end{equation}
Our determinations appear to be in good agreement with the results in
Ref.~\cite{Cea:2012qw} which were obtained using the connected correlator built
with the Wilson loop, and the cooling procedure. Type-I superconductivity of the
SU(3) vacuum is confirmed and agreement is found also with
Ref.~\cite{Shibata:2012ae,*Cardoso:2013lla}.

\section*{Acknowledgments}\vspace{-0.25cm}
Simulations have been performed on the BC$^2$S cluster in Bari.
We also acknowledge support from the INFN - SUMA project.\vspace{-0.25cm}

\small{
}

\end{document}